\documentclass[aps,prl,twocolumn,groupedaddress]{revtex4}

\newcommand{\bb}{\begin{equation}}
\newcommand{\ee}{\end{equation}}
\newcommand{\ba}{\begin{eqnarray*}}
\newcommand{\ea}{\end{eqnarray*}}
\newcommand{\rhor}{\rho({\bf r})}
\newcommand{\dd}{{\rm d}}
\newcommand{\rr}{{\mathbf r}}
\newcommand{\dr}{{\rm d}{\bf r}}

\usepackage{graphicx}
\usepackage{dcolumn}
\usepackage{bm}
\usepackage{longtable}

\begin{document}


\title{Microscopic Density Functional Theory for Dendrimers}

\author{Alexandr \surname{Malijevsk\'y\footnote{Electronic address: a.malijevsky@imperial.ac.uk}}}

\affiliation{%
Department of Chemical Engineering, Imperial College London, South
Kensington Campus, London SW7 2AZ, United Kingdom and
E. H{\'a}la Laboratory of Thermodynamics, Institute of Chemical Process Fundamentals
      of the ASCR, 165 02 Prague 6, Czech Republic}%


\date{\today}
\begin{abstract}
Density functional theory for a simple model of dendrimers is
proposed. The theory is based on fundamental measure theory which
accounts for the hard-sphere repulsion of the segments and on the
Wertheim first-order perturbation theory for the correlations due
to connectivity. Set of the recurrence formulae for the ideal
chain contribution involving simple integrals is derived. By using
perturbation theory dispersion forces can be easily included.
\end{abstract}

\pacs{Valid PACS appear here}

\maketitle

%
Dendrimers (also known as arborols or cascade molecules) are
repeatedly branched monodisperse compounds with a fractal-like
structure possessing a high degree of symmetry. The first
dendritic structure was constructed by V\"{o}gtle et al. in 1978
\cite{vogtle} using a repetitive synthesis strategy (divergent
synthesis) such that one new molecular layer (generation) is
created in each reaction cycle. Almost immediately dendrimers
attracted increasing attention due to their unique structure
implying some unique properties compared to traditional linear
chains. With increasing number of generations the number of
chain-ends increases exponentially and that is why dendrimers
adopt a compact globular shape. As a consequence the
dendrimers' solubility is driven by the nature of surface groups
only \cite{klajnert}. Moreover, the presence of internal cavities
enables encapsulation of small guest molecules \cite{jansen}. Such
properties predestinate dendrimers for a variety of possible
technological application. They are being developed for use in
fields such as catalysis, magnetic resonance imaging, drug
delivery, coatings, electronics or cancer therapy
\cite{hoover,hu,haba,boubbou}.

In terms of theoretical study and in particular of physical model
used, there are in principle two perspectives in treating such
complex structures. First, one can use so-called coarse graining
technique treating molecules as spherically symmetric objects
interacting via some effective soft potential, i.e. the degrees of
freedom of monomers composing the polymer are integrated out.
Perhaps the first who proposed such a strategy was Flory
\cite{flory}, but it has become popular only in recent times
\cite{louis, likos}. The coarse-grained approach was applied for
dendrimers by G\"{o}tze et. al \cite{gotze} who approximated the
interaction between two dendrimers in solution by an appropriate
Gaussian function. Even though the coarse-graining method is
attractive in simplifying the description of a given complex
structure substantially, it inevitably loses some information
about the intrinsic property of the molecules. Moreover, the
radial symmetry of an effective potential is more justified in a
bulk phase rather than e.g. in the vicinity of a wall. For these
reasons, a second approach, treating the complex molecular systems
on an atomistic level, would seem to be to superior. Microscopic
density functional (DF) theory provides a versatile and powerful
tool to represent the microscopic structures and interfacial
phenomena of polyatomic fluids under a variety of situations. 
Woodward developed a theory that combines
weighted density approximation, known from theories of
simple fluids, with single-chain Monte Carlo simulations \cite{woodward}. An
alternative DFT of inhomogeneous polymer solutions was
formulated by Forsman et al. \cite{forsman}. Their
theory is based on the free energy functional resulting from
the generalized Flory equation of state.
However, an approach due to Yu and Wu \cite{yu} which incorporates Wertheim's
perturbation theory for a bulk fluid \cite{wertheim, saft} into
the non-local DF framework proposed by Rosenfeld \cite{rosenfeld} revealed to be particularly appealing.
In the spirit of the Rosenfeld fundamental measure theory (FMT),
Yu and Wu constructed a non-local functional accounting for the
chain connectivity. Such an approach has proved to be both
quantitatively accurate and computationally convenient and
eventually has been extended for a variety of models of e.g.
cyclic polyatomic fluids \cite{cyclic}, block copolymers
\cite{bc}, star polymers \cite{star}, polydisperse polymers
\cite{polydisperse} or for brush-like structures \cite{brush}.
Recently, a so-called hybrid approach for the 
structure of dendrimers has been proposed \cite{cao_dend} 
in the spirit of Ref. [12].
In this paper, a full density functional
is derived and used to represent hard dendrimers confined between
two hard walls.

In the following, by the term ``dendrimer'' will be taken to mean
the special case of a tree structure where each segment apart from
 the terminating ones has the same number of bonds, three at minimum.
Due to the high level of symmetry the system can be characterized
by two parameters:
 i) $f$, number of arms, i.e. bonds outcoming from each (except the terminating)
 segment;
 ii) $M$, the number of generations, i.e. number of segments contained in a chain connecting the central
 and terminating segment minus one.
Dendrimers with $f$ arms and $M$ generations will be abbreviated
by $D(f,M)$.

Each dendrimer contains a central a segment, which is by
definition segment of generation 0. Segments of the $i$th
generation are connected to the central segment by a chain of
$i+1$ segments.
The number of segments of $i$th generation is
$g_i=f\cdot(f-1)^{i-1},\;i\geq1$ and the total number of segments
is $N=f\frac{1-(f-1)^M}{2-f}+1$.

Segment positions are labeled by two indexes; the subscript,
$i=0..M$, specifying the generation, and the superscript,
$j=1..g(i)$, specifying the position in a given generation. The
latter can be set in a clockwise order (in a two-dimensional
projection) such that segments $\rr_i^1,\ldots,\rr_i^{f-1}$ are
connected to the segment $\rr_{i-1}^1$. Position of a whole
dendrimer can be expressed by a vector
 ${\bf R}=\prod_{i=0}^{M}\prod_{j=1}^{g(i)}\rr_i^j$.

The model under interest will be represented by tangentially
connected hard spheres of diameter $\sigma$, each interacting via
potential $\psi$ with an external field. The grand potential
functional of such a system can be expressed as \cite{yu}

\bb
\beta\Omega[\rho_N({\bf R})]=\beta F_{\rm id}[\rho_N({\bf
R})]+\beta F_{\rm ex}+\int[\Psi({\bf R})-\mu]\rho_N({\bf
R})\dd{\bf R}\,,
\ee
where
\bb
\beta F_{\rm id}=\int\dd{\bf R}\rho_N({\bf R})[\log\rho_N({\bf R})-1]+\beta\int\dd{\bf R}\rho_N({\bf R}) V_b({\bf R})
\ee
is the contribution corresponding to the system of ideal chains
that interact only through bounding potential, $V_b({\bf R})$, and the excess part that takes into account correlations between
nonbonded segments
\bb
\beta F_{\rm ex}=\int\dr\left\{\Phi^{\rm hs}[n_\alpha(\rr)]+\Phi^{\rm c}[n_\alpha(\rr)]\right\}
\ee
is split into the hard-sphere contribution $\Phi^{\rm hs}$ and the contribution due to the chain connectivity $\Phi^{\rm c}$.

$\rho_N({\bf R})$ is the dendrimer density and $\Psi({\bf
R})=\sum_{i=0}^{M}\sum_{j=1}^{g(i)}\psi(\rr_i^j)$. Further,
$V_b({\bf R})$ is a sum of bounding potentials between the
neighboring segments creating the dendrimer structure,
\bb
\exp[-\beta V_b\left({\bf R}\right)]=\prod_{i=0}^{M-1}\prod_{j=1}^{g(i)}\prod_{k=-\delta_{i0}}^{f-2}\frac{\delta(|\rr_i^j-\rr_{i+1}^{j(f-1)-k}|-\sigma)}{4\pi\sigma^2}\,.
\ee

Free energy densities  $\Phi^{\rm
hs}[n_\alpha(\rr)]$ and $\Phi^{\rm
c}[n_\alpha(\rr)]$ are functions of four scalar and two vector
weighted densities $\{n_\alpha(\rr)\}$ \cite{yu,rosenfeld}. For
the hard-sphere contribution, $\Phi^{\rm hs}[n_\alpha(\rr)]$, the
so-called White-Bear approach (or modified FMT)
have been used, see Refs. \cite{yu, wb} for the explicit formulae.

The free energy density due to indirect chain connectivity,
$\Phi^{\rm c}[n_\alpha(\rr)]$, was obtained  as a generalization
of Wertheim's first-order perturbation theory for a bulk fluid
\cite{wertheim, saft}  for inhomogeneous systems within the
nonlocal DF framework \cite{yu}

\bb \Phi^{\rm c}[n_\alpha(\rr)]=\frac{1-N}{N}n_0\zeta\ln[g_{\rm
HS}(\sigma,\{n_\alpha\})]\,, \ee where $\zeta=1-{\bf
n}_{2}\cdot{\bf n}_{2}/(n_2)^2$, and the contact value of the
hard-sphere pair correlation function, $g_{\rm
HS}(\sigma,\{n_\alpha\})$, is obtained from the Carnahan-Starling
equation of state. The important feature of this approach is that
the problem is formulated on the level of average segment density,
$\rhor$, which is related to the density of a whole dendrimer
$\rho_N({\bf R})$ via


\bb
\rhor=\sum_{i=0}^{M}\sum_{j=1}^{g(i)}\rho_i^j(\rr)=\sum_{i=0}^{M}\sum_{j=1}^{g(i)}\int\dd{\bf
R}\delta(\rr-\rr_i^j)\rho_N({\bf R})
\ee
where $\rho_i^j(\rr)$ is the density distribution of an individual
segment.

A minimization of the grand potential functional with respect to
the density distributions gives rise to a set of the following
Euler-Lagrange equations: \bb
\rho_i^j(\rr)=\exp(\beta\mu)\int\dd{\bf
R}\delta\left(\rr-\rr_i^j\right)\exp\left[-\beta V_b({\bf
R})\right]\gamma(\rr) \ee and \bb
\rho(\rr)=\exp(\beta\mu)\int\dd{\bf
R}\sum_{i=0}^{M}\sum_{j=1}^{g(i)}\delta\left(\rr-\rr_i^j\right)\exp\left[-\beta
V_b({\bf R})\right]\gamma(\rr)\,. \ee
 Due to symmetry,
$\rho_i^j(\rr)$ depends on its generation number only, so that the
upper index will be omitted. Function $\gamma(\rr)$ is defined as

\bb
\gamma(\rr)=\exp\left\{-\beta\prod_{i=0}^{M}\prod_{j=1}^{g(i)}\left[\frac{\delta F_{\rm ex}}{\delta\rho_i^j\left(\rr\right)}+\psi\left(\rr_i^j\right)\right]\right\}\,.
\ee

Specifically, the segment density distributions have been
calculated for hard-sphere dendrimers confined by two plane hard
walls placed a distance $H$ apart, i.e. the external field
interacts with each segment with a potential

\begin{equation}
\psi(z)=\left\{\begin{array}{ll}\infty&z<\sigma/2 \;{\rm or}\; z>H-\sigma/2\nonumber\\
0&{\rm otherwise} \end{array}\right. 
\end{equation}
For this system $\rho_i^j(\rr)=\rho_i^j(z)$, and the the Euler-Lagrange
equations have much simpler forms: \bb
\rho_0(z)=\exp(\beta\mu)\gamma(z)(G_M(z))^f \ee and \bb
\rho_i(z)=\exp(\beta\mu)\gamma(z)(G_{M-i}(z))^{f-1}\tilde{G}_i(z)\;i\geq1\,.
\ee The functions $G_i(z)$ and $\tilde{G}_i(z)$ are defined by the
following recurrence relations \bb G_i(z)=\int\dd
z'\gamma(z')(G_{i-1}(z'))^{f-1}\frac{\theta(\sigma-|z-z'|)}{2\sigma}\;i\geq1
\ee and \bb \tilde{G}_i(z)=\int\dd
z'\gamma(z')\tilde{G}_{i-1}(z')(G_{M-i+1}(z'))^{f-2}\frac{\theta(\sigma-|z-z'|)}{2\sigma}\;i\geq2\,,
\ee with $G_0(z)=1$ and

$$\tilde{G}_1(z)=\int\dd z'\gamma(z')(G_M(z'))^{f-1}\frac{\theta(\sigma-|z-z'|)}{2\sigma}.$$

\begin{figure} [tbp]
\includegraphics[clip,width=8cm]{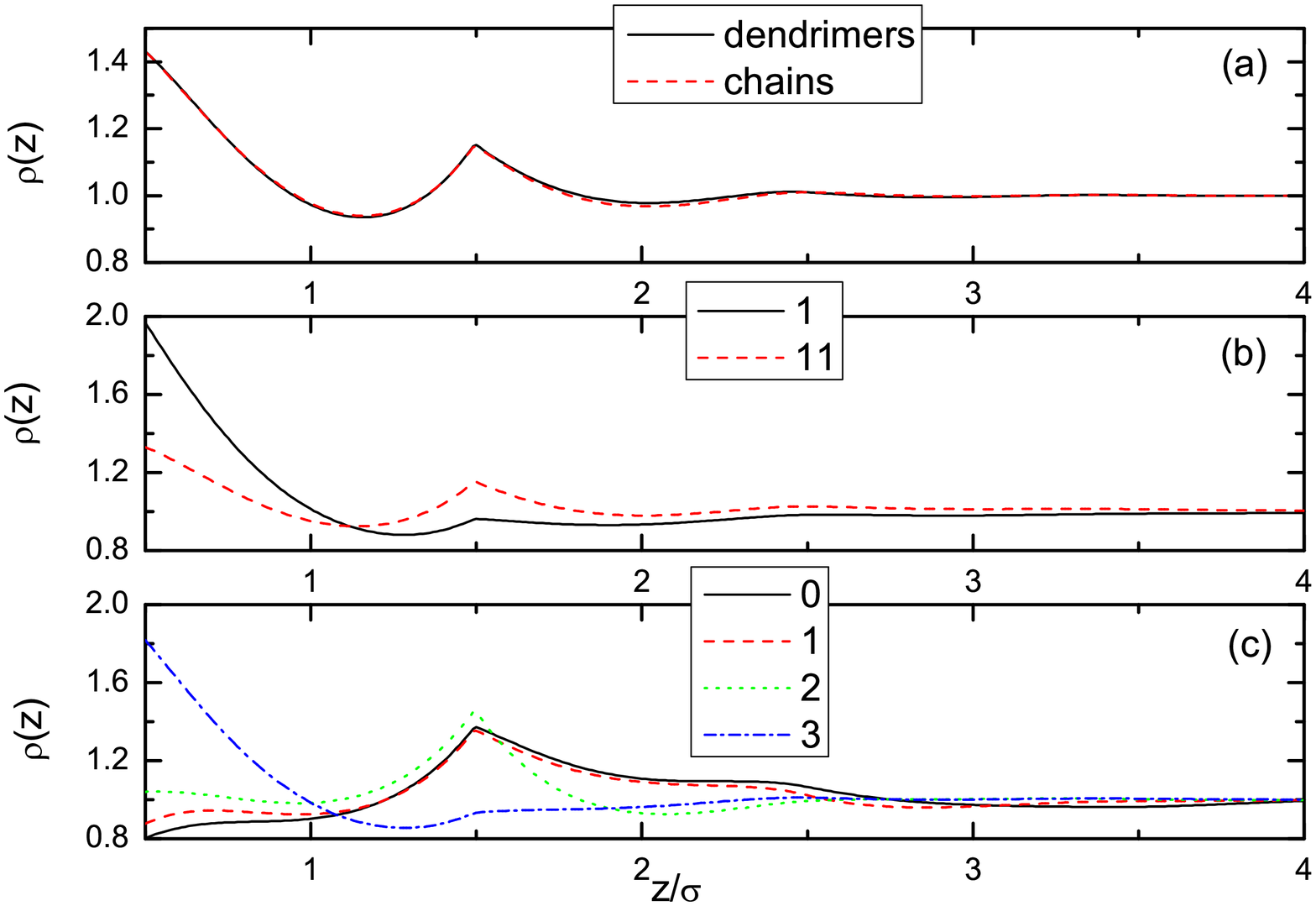}
\caption{(a) The average segment density of third generation
dendrimers, $D(3,3)$, and linear chains, of $N=22$ segments. (b)
Segment density profiles of the first and eleventh segments of
linear chains, of $N=22$ segments. (c) Segment density profiles of
the zeroth, first, second, and third generations of the dendrimer
$D(3,3)$. The bulk segment density is $\rho_b^*=0.5$. The hard
walls are separated by $H=10\sigma$.}
\end{figure}

In Fig.~1 comparison of the density profiles for third generation
$D(3,3)$ with the profiles of linear chain polymers comprising the
same number of tangent segments, i.e. $N=22$, is shown. The latter
are obtained from the theory of Yu and Wu \cite{yu}. The
calculations have been performed for an average bulk segment density of
$\rho_b^*=\rho_b\sigma^3=0.5$. It is evident from the upper panel
that the average segment densities of dendrimers and chains are
very similar for such a density. Both profiles exhibit oscillation
characteristic with adsorption on the wall. The system is thus in
the regime where the structure of the fluid is dominated by
excluded volume effects. In such a case the specific architecture
of the molecules plays a less important role and the system as a
whole behaves like a hard-sphere system. It is interesting to note
the identical contact densities, $\rho(\sigma/2)$, which reflects
an equality of bulk pressures of both systems according to the sum
rule \cite{Henderson}. This is because the systems are treated in
the TPT1 approximation where only number of bonds (not their
topology), which are same for both models, are taken into account.

Although the average segment densities are nearly identical, there
are large differences in the densities of individual segments. In
Fig.~1b the density profiles of the terminal and middle segments
of chains are plotted. They both exhibit qualitatively similar
behaviour to the average segment density, with higher value of
contact density for the terminal segment. This is because of a
smaller loss of orientational entropy if the terminal rather
than the middle segment is at the contact with the wall in the case of chains
and because of excluded volume interactions from the outer segments in the case of dendrimers. The
density profiles of the remaining segments smoothly interpolate
between these two curves. The behaviour observed for dendrimers is
very different, see Fig.~1c. In this case only the terminal
segments are in a regime where excluded volume effect dominates
whereas all other segments exhibit surface depletion. Clearly, the
terminal segments can be adsorb on to the wall more easily then
those of lower generations. Because the number of segments of the
highest generation is more than half of the total number of
segments the adsorption for the average segment density persists.
In all cases a cusp in the density profiles a distance $\sigma$
from the sphere-wall contact position which reflects harshness of the fluid-fluid and the
fluid-wall interaction.

\begin{figure} [tbp]
\includegraphics[clip,width=8cm]{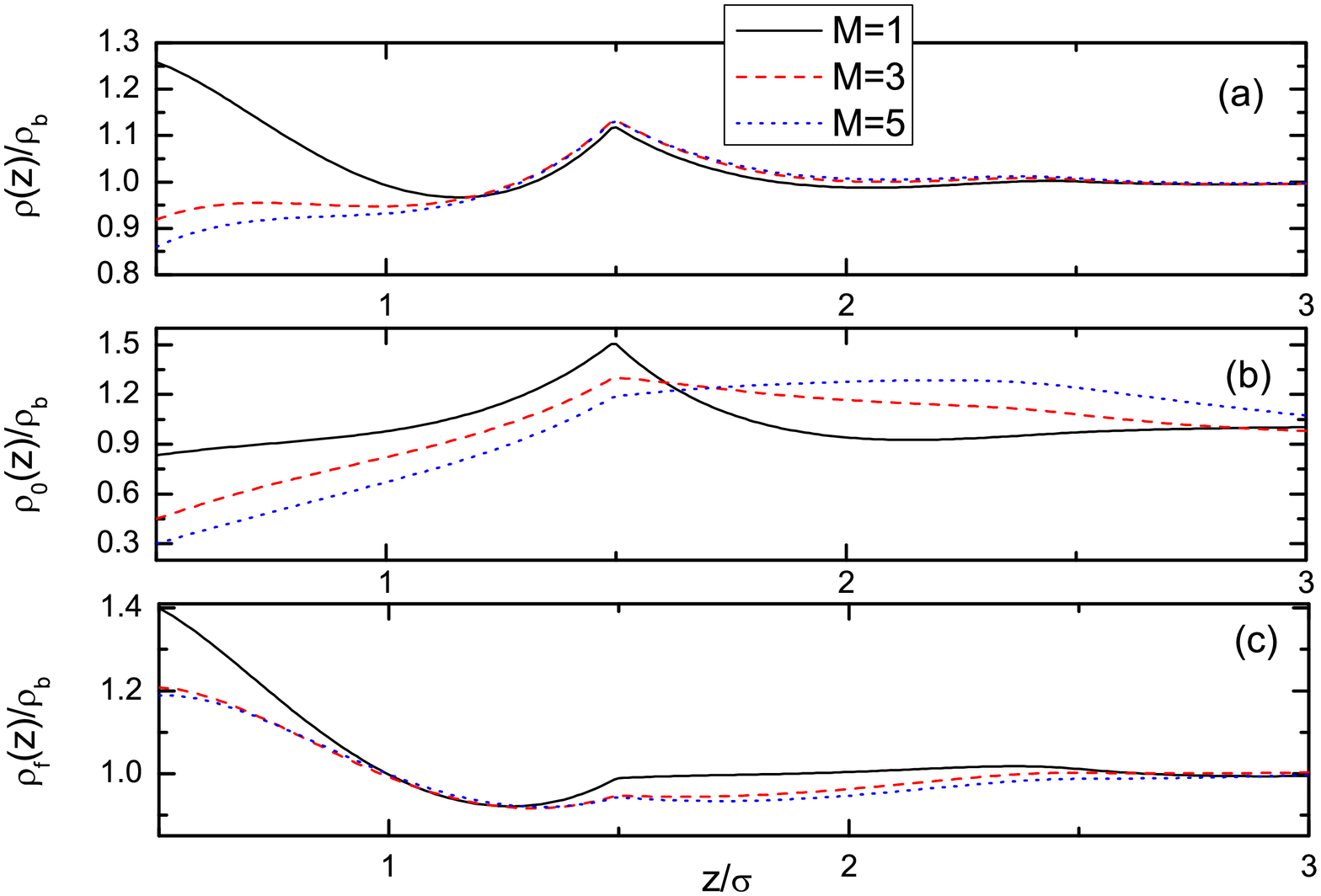}
\caption{Density profiles of dendrimers $D(3,1)$, $D(3,3)$, and
$D(3,5)$ for $H=10\sigma$ and $\rho_b^*=0.4$. (a) The average segment
density; (b) Segment density of cores (segments of zeroth
generation); (c)  Segment density of terminal segments.}
\end{figure}

The impact of the generation number on a structure of the fluid is
examined in Fig 2. The calculations are carried out for $D(3,1)$,
$D(3,3)$, and $D(3,5)$ for a bulk density $\rho_b^*=0.4$. Now
significant differences  for different architectures are apparent
in the average density profiles, see Fig.~2a. We observe a
transition from surface adsorption for $D(3,1)$ (the simplest star
polymer) to depletion for $D(3,3)$ and $D(3,5)$. Interestingly for
the third generation dendrimer  $D(3,3)$ the effect of depletion
and adsorption almost compensate each other. The density profiles
of the zeroth generation segments (the cores) and the terminal
segment are compared in Figs. 2b and 2c, respectively. Whereas
in all three cases the zeroth segments exhibit depletion,
adsorption is always found for terminal segments. Larger
differences are apparent for the central segments, particularly in
the slopes of the profiles beyond the cusps which change from
negative to positive with increasing $M$, being close to zero for
$D(3,3)$.

\begin{figure} [tbp]
\includegraphics[width=8cm]{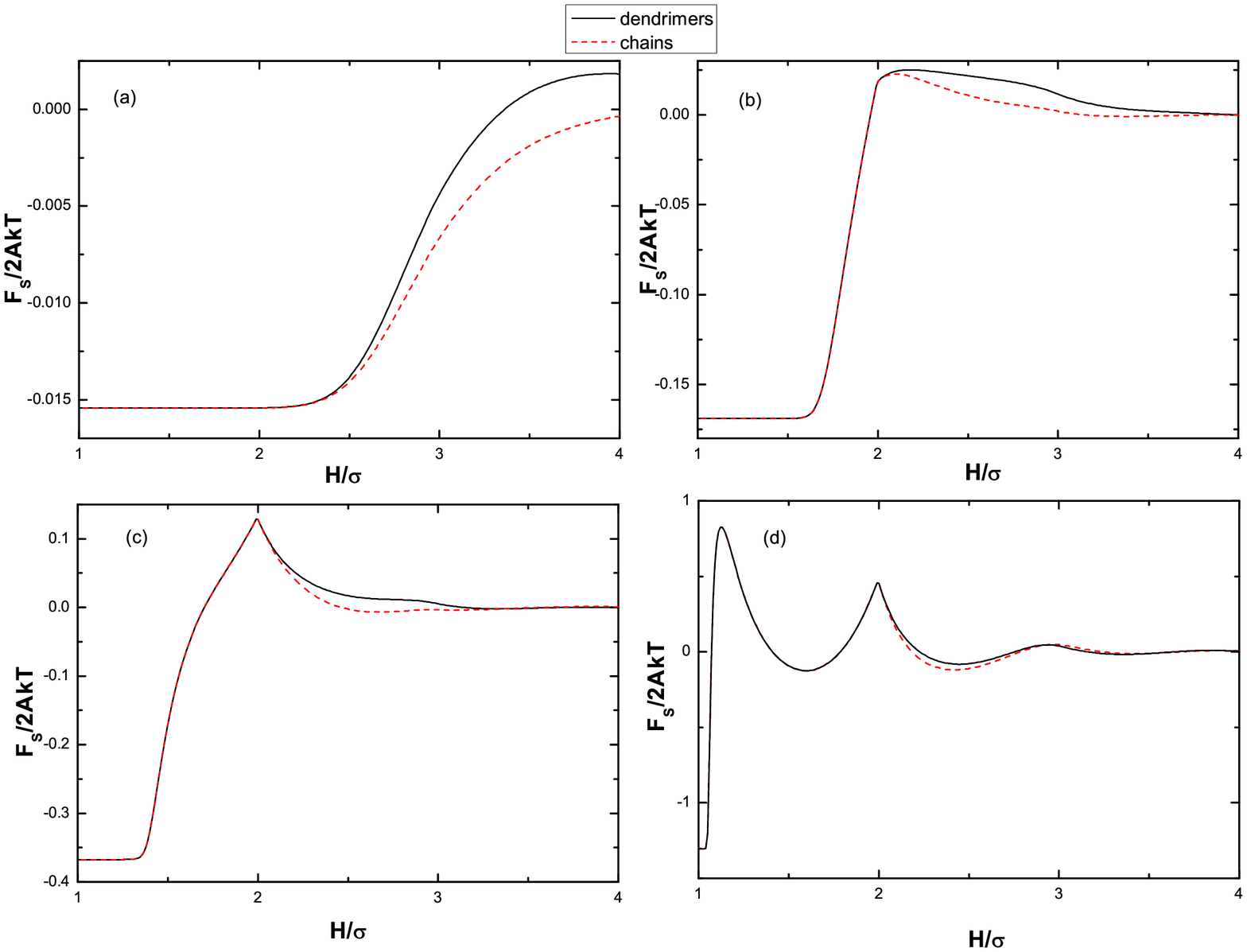}
\caption{Solvation forces between two hard walls separated by
third generation dendrimers $D(3,3)$ (continuous curves) and
linear chains of $N=22$ segments, (dashed curves). The average bulk
segment densities are (a) $\rho_b=0.1$, (b) $\rho_b=0.3$, (c)
$\rho_b=0.4$, (d) $\rho_b=0.6$}
\end{figure}

In a final analysis the solvation forces between the two hard
walls are calculated for two models, $D(3,3)$ and the equivalent
linear chains of $N=22$ tangent segments. According to the sum
rule \cite{Henderson}, the solvation force $F_S$ per unit area $A$
is related to the average contact density through $
F_S/(2Ak_bT)=\rho(0)-\rho_\infty(0)\,,$ where $\rho_\infty(0)$ is
the average contact density for infinite separation. In Figure 3
the four regimes corresponding to bulk densities $\rho_b=0.1 ,0.3,
0.4$, and $0.6$ that the solvation forces can obey for the two
models under consideration are presented. At the lowest density,
Fig.~3a, the depletion forces dominate, so that the solvation
force is attractive for small separations. At larger separations
the attraction decays monotonically to zero in the case of linear
hard-sphere chains whereas for dendrimers $F_S$ first changes sign
and eventually converges to zero. For intermediate densities,
Fig.~3b and Fig.~3c, both solvation force profiles exhibit a
maximum which is smooth for $\rho_b=0.3$ but which for the higher
density of $\rho_b=0.4$ becomes a cusp. At the highest density of
$\rho_b=0.6$, Fig.~3d, the specific topology of molecules becomes
irrelevant and both solvation force profiles have very similar
oscillatory characteristics of hard-sphere systems.

In this work a density functional theory for a primitive model of
dendrimers is proposed.
A compact recursive formulae is derived
involving both intra- and inter-molecular forces taking the form
of simple integrals which greatly facilitates the numerical
calculations. This is the first step in a theoretical treatment of
more realistic models of dendrimers for which the  theory can be straightforwardly extended by using of a
perturbation technique. It will enable to study various interesting problems; for instance, one of the controversy is whether dendrimers of higher generations adopt a membrane-like surface \cite{caminati} or exhibit rather homogeneous
segmental density due to back-folding effects \cite{likos01}. Behavior of dendrimers in concentrated solutions is also of an interest. 
The quantitative agreement of DF theory for star polymers (the simplest case of dendrimers) derived on a similar basis \cite{star} with Monte-Carlo data gives one confidence that the theory developed in the present work provides an accurate representation of dendrimeric systems. However, only more detailed numerical tests for corresponding models will reveal to what extent the proposed theory for dendrimers is appropriate, a matter of current research.



\end{document}